\documentclass[aps,preprint,showkeys,superscriptaddress, groupedaddress, nofootinbib]{revtex4}  
\usepackage{latexsym}
\usepackage{amsmath}
\usepackage{booktabs}
\usepackage[utf8x]{inputenc}
\usepackage[T1]{fontenc}
\usepackage{subdepth}
\usepackage{color}
\usepackage{soul}

\begin{document}
\def\calr{{\cal R}}
\def\GB{{\hat{\cal{G}}}}
\newcommand{\RR}{(*R*)}
\def\half{\textstyle{1\over2}}
\def\third{\textstyle{1\over3}}
\def\quarter{\textstyle{1\over4}}
\def\nn{\nonumber}
\def\Vg{V_{george}}
\def\Vr{V_{ringo}}
\def\Vj{V_{john}}
\def\Vp{V_{paul}}
\def\tap{{}^+\tau}
\def\tam{{}^-\tau}
G. P\'erez-Cu\'ellar $^{\ddagger}$ and M. Sabido $^{\ddagger}$*
\title{On planetary orbits, ungravity and entropic gravity}
\author{G. P\'erez-Cu\'ellar$^1$}
\email{perezcg2015@licifug.ugto.mx}
\author{M. Sabido$^{2}$}
\email{msabido@fisica.ugto.mx}
\affiliation{Departamento  de F\'{\i}sica de la Universidad de Guanajuato, A.P. E-143, C.P. 37150, Le\'on, Guanajuato, M\'exico.
 }%
\begin{abstract}
In previous works, entropic gravity and ungravity have been considered as possible solutions to the dark energy and dark matter problems. To test the viability of these models, modifications to planetary orbits are calculated for ungravity and different models of entropic gravity. Using the gravitational sector of  unparticles, an equation for the contribution to the effect of orbital precession is obtained. We conclude that the estimated values for the ungravity parameters from planetary orbits are inconsistent with the values needed for the cosmological constant. The same ideas are explored for entropic gravity arising  from a modified entropy--area relationship.
 \end{abstract}

\maketitle

\section{Introduction}
General Relativity (GR) has been a successful theory since it was postulated by Einstein in 1915. The~first solution of the Einstein field equations by Schwarzschild and the first experimental proof of the effects described by this theory, such as the light deflection observation by Eddington, came soon after it was published. Since then, GR has been successfully tested in several phenomena in nature, from~simple low-energy systems, such as the orbital precession in the solar system, to~more complex high-energy systems, {such} as neutron stars or black holes.
However, some effects remain unexplained, such as the~case of dark matter and dark energy~\cite{dark,energy}, which we have concluded to be essential components of the universe, but~until now we have been unable to describe the mechanisms involved in observations or find an appropriate frame to describe them.  Also, the~recent experimental confirmation  {of} the existence of black holes~\cite{event} forces us to understand the most fundamental aspects of gravity. Moreover, black holes are gravitational systems in which quantum effects can be important, and~due to our ignorance of quantum gravity, alternative approaches must be considered. For~example, in~the semiclassical approach, some macroscopic effects with information about the hidden quantum degrees of freedom exist. Such proposals were made by Bekenstein, Hawking and Unruh in the 1960s. Then, following the analogy between gravity and thermodynamics, Jacobson wrote Einstein's equations as an equation of state~\cite{jacobson}.
{The consideration of} gravity  as an emergent phenomenon allows the use of the statistical mechanics framework to study this interaction. This idea was revived in~\cite{verlinde_2011}, where it was proposed that Newtonian gravity is an entropic force,  analogous to emergent forces  in the study of polymers. The~motivation is based on the idea of holography and its relation to black holes. In~this formulation, one can propose modifications to Newtonian gravity by analyzing modifications to the {entropy--area} relationship. This approach to gravity has been  used to study several gravitational phenomena in~connection to anomalous galactic rotation curves~\cite{isaac}, late time acceleration of the universe and dark energy~\cite{isaac_2,isaac_3}, {or} black hole quasinormal modes~\cite{pinki}; different modifications to the Hawking--Bekenstein  entropy--area relationship are used to~modify either the Newtonian equation for gravity or the Friedmann equations and~therefore study the effects and  modifications to~GR.

In a  completely different context from entropic gravity and using different physical principles, one can derive another type of modified gravity. We can start by considering higher-energy extensions to the Standard Model. That is the case of unparticles, a~scale-invariant hidden sector of the Standard Model proposed in~\cite{georgi}. The~components of this sector, called unparticles (unlike particles), have a continuous mass spectrum and a characteristic energy for interactions with SM particles~\cite{Nikolic:2008ax}. In~the context of gravity, one can understand ungravity as the result of ungraviton interaction{s}. To~introduce the effects of unparticle physics  {on} gravity, one adds an unparticle term to the Hilbert--Einstein action. In~\cite{Gaete:2010sp}, the~authors studied the ungravity counterpart of the Schwarzschild black hole. Moreover, using ungravity's temperature and entropy, the~ungravity sector effects have been studied in {cosmology, allowing} to relate ungravity parameters with the cosmological constant value~\cite{Diaz-Barron:2019uzd}. In~order to see if this result is consistent with other gravitational scenarios, we can compare it to planetary motion.
Therefore, the~main goal of this paper is to determine if  ungravity and entropic gravity  are consistent with~observations.

The paper is arranged as follows: In Section~\ref{sec:ungrav}, we obtain orbital precession from a modified Schwarzschild ungravity metric. In~Section~\ref{sec:entropic}, a brief review of Newtonian entropic gravity is presented, and modifications to orbital precession are calculated. Corrections are obtained from a generalized entropy--area relationship. As~in the ungravity case, the~corrections are evaluated using solar system data. Finally, Section~\ref{sec:discussion} is devoted to discussion and final remarks.
\section{Ungravity Contributions to the Orbital~Precession}
\label{sec:ungrav}
In this section, we consider the unparticle  generalization. This theory is known as ungravity and  is constructed by considering ungraviton interactions. 
The action is constructed as the sum of the Einstein--Hilbert action, the~matter action  {and the}  effective action $S_U$ for the ungravitons. The~ungravity action~\cite{gaete} is given by 
\begin{equation}
S_U=\frac{1}{2k^2}\int d^4x\sqrt{g}R\left [ 1+\frac{A_{d_U}}{(2d_U-1)\sin{\pi d_U}}\frac{\kappa_*^2}{\kappa^2}\left (\frac{-D^2}{\Lambda_U^2}\right)^{1-d_U}\right ]^{-1},
\end{equation}
where $D^2$ is the D'Alemberteian, $\kappa_*$ is the ungravitational Newton constant, $\kappa=16\pi G_N$  and $A_{d_U}$ is a constant.\footnote{This constant  depends on the ungravity parameter $d_U$ and  gamma functions $A_{d_U}=\frac{16\pi^{5/2}}{(2\pi)^{2d_U}}\frac{\Gamma(d_U+1/2)}{\Gamma(d_U-1)\Gamma(2d_U)}$.} 
 The~modified Einstein equations are
\begin{equation}
G_{\mu\nu}=\kappa^2T^\mu_{~~\nu}+\kappa_*^2\frac{A_{d_U}}{\sin{\pi d_U}}.
\end{equation}
Solving 
 for the Schwarzschild black hole, one obtains the ungravity Schwarzschild metric~\cite{Gaete:2010sp}
\begin{equation}
   g_{r r}^{-1}=-g_{00}=1-\frac{2 M G_N}{r}\left[1+\kappa_*^2 \frac{A_{d_U}}{\sin \left(\pi d_U\right)} \frac{M(r)}{2 M G_N}\right] 
\end{equation}
{where}
\begin{equation}
    M(r)=\frac{2^{2 d_U-2}}{4 \pi^{1 / 2}} \frac{\Gamma\left(d_U-1 / 2\right)}{\Gamma\left(2-d_U\right)} M \Lambda_U^{2-2 d_U}\left(\frac{1}{r}\right)^{2 d_U-2}.
\end{equation}
This leads to the following modified metric:
\begin{equation}
    g_{rr}^{-1} = -g_{00} = 1 -\frac{2G_NM}{r}\left[1 - \left(\frac{R_G}{r}\right)^{2d_u - 2}\right],
    \label{eq:mod0r}
\end{equation}
where 
$R_G$ is related to the ungravity  parameters~\cite{georgi,Gaete:2010sp} as

\begin{equation}\label{eq:Rg}
      R_G  = \frac{1}{ \pi \Lambda_u}\left(\frac{M_{pl}}{2 M_u}\right)^{1/(d_u - 1)}\left(\frac{1}{2\pi}\frac{\Gamma(d_u + \frac{1}{2})\Gamma(d_u - \frac{1}{2})}{\Gamma(2d_u)}\right)^{1/(2du - 2)}. 
\end{equation}

This term can be understood as the length scale at which unparticle effects will be relevant and is a free parameter of the model, together with $\Lambda_u$, the~characteristic energy of the model; $M_u$, the~characteristic mass; $d_u$, the dimension of the extra operators in the action; and~ Planck's mass $M_{pl}$.

Let us now calculate the orbital precession using the ungravity Schwarzschild metric. Considering a particle  near a spherically symmetric gravitational field,  the geodesic equation~is
\begin{eqnarray}\label{eq:geodesicsK}
        2K &=&  \left( 1 - \dfrac{2GM}{r}\left[1 - \left(\dfrac{R_G}{r}\right)^{2d_u - 2}\right]\right)\Dot{t}^2\nonumber\\
         &-&  \left(1 - \dfrac{2GM}{r}\left[1 - \left(\dfrac{R_G}{r}\right)^{2d_u - 2}\right]\right)^{-1}\Dot{r}^2 - r^2\Dot{\theta}^2 - r^2\sin^2\theta\Dot{\phi}^2,
\end{eqnarray}
where $M$ is the gravitational mass, {and} the  dots denote a derivative with respect to the proper time (also, we set $c = 1$). The~above equation is solved taking $2K = 1$, as~we are considering time-like geodesics. The~Euler--Lagrange equation leads to the following conserved quantities:
\begin{equation}
    q \equiv \left( 1 - \dfrac{2GM}{r}\left[1 - \left(\dfrac{R_G}{r}\right)^{2d_u - 2}\right]\right)\Dot{t},\qquad  h \equiv r^2\sin^2\theta\Dot{\phi}.
    \label{eq:conservedqs}
\end{equation} 
When considering planar motion $\theta_0 = \frac{\pi}{2}$, the~conserved quantity $h$ can be identified with angular momentum per unit mass, in~analogy with the usual Kepler problem. An~equation of motion is obtained in terms of the constant $q$, which {is related to} the conservation of energy. 
Using the change of variable $u = 1/r$ in this $\theta_0$ plane, we obtain 
\begin{equation}
    \left(\frac{du}{d\phi}\right)^2 + u^2 = \frac{q^2 - 1}{h^2} + 2GM\left(\frac{u}{h^2} + u^3\right)\left[1 - (uR_G)^{2d_u - 2}\right].
\end{equation}
Finally, differentiation with respect to $\phi$ together with  {$\mu \equiv GM$} gives rise to a modified  Binet equation,
\begin{equation}\label{eq:diffecRGorbits}
    \frac{d^2u}{d\phi^2} + u = \frac{\mu}{h^2}\left[1 + 3u^2h^2 
    - (2d_u - 1)(uR_G)^{2d_u -2}-(2d_u + 1)h^2R_G^{2d_u - 2}u^{2d_u}\right].
\end{equation}
Using the parameter $\varepsilon \equiv 3\mu^2/h^2$ to solve perturbatively, we propose the  solution\footnote{The assumption  is that the ungravity contribution is of the same order as for GR, for~the purpose of  comparing with experimental data. This  is consistent with  $1 < d_U < 2$ \cite{georgi2007}, which are usually the stated values for this parameter.}  $u = u_0$ + $\varepsilon$ $u_1$.  The~zeroth-order equation $u''_0 + u_0 = \mu/h^2$ has the usual conic section solution, and~the first-order differential \mbox{equation is}
\begin{equation}
\begin{split}
     u''_1 + u_1 =& \frac{\mu}{h^2}(1 + e\cos\phi)^2 - \frac{\mu^{2d_u -3}}{3h^{4d_u -4}}(2d_u - 1)R_G^{2d_u -2}(1 + e\cos\phi)^{2d_u -2}\\
     &- \frac{\mu^{2d_u -1}}{3h^{4d_u -2}}(2d_u + 1)R_G^{2d_u -2}(1 + e\cos\phi)^{2d_u},
     \label{eq:diffecRGu1}
\end{split}
\end{equation}
where $u_1''$ denotes derivatives with respect to $\phi$, and $e$ denotes the orbital eccentricity. The~first term in the right-hand side can be identified as the usual GR contribution to the Kepler problem. After~setting\footnote{This value for $\beta$ is phenomenologically relevant. It gives an entropy proportional to a volumetric contribution, and in the cosmological scenario gives an effective cosmological constant~\cite{Gaete:2010sp,Diaz-Barron:2019uzd}.} $d_u = \frac32$, we solve the equation above by considering only linear contributions in $\phi$. Then, the~solution to the first-order differential equation is
\begin{equation}
    u_1 = \left(\frac{\mu}{h^2} - \frac{R_G}{3h^2} - \frac{2R_G\mu^2}{h^4} - \frac{R_G\mu^2}{2h^4}e^2\right)e\phi \sin\phi.
\end{equation}

Finally, the~complete solution $u = u_0 + \varepsilon u_1$ is
\begin{equation}
    u = \frac{\mu}{h^2}\{1 + e\cos[\phi(1 - \delta)]\},
\end{equation}
where
\begin{equation}
    \delta \equiv \dfrac{3\mu^2}{h^2} - \dfrac{R_G \mu}{h^2} -  \dfrac{6R_G\mu^3}{h^4} - \dfrac{3R_G\mu^3e^2}{2h^4}\ll 1.
\end{equation}

The orbital precession is calculated by taking a complete period $\phi_T(1 - \delta) = 2\pi$ such that the extra contribution represents the precessed angle. In~this case, {the} ungravity contribution to the precession on each revolution is
\begin{equation}
    \vert\Delta\phi_u\vert =2\pi\left(\frac{R_G \mu}{h^2} + \frac{6R_G\mu^3}{h^4} + \frac{3R_G\mu^3e^2}{2h^4}\right),
    \label{eq:deltaP}
\end{equation}
which can be rewritten in terms of astronomical variables $h^2 = GMa(1-e^2)$
\begin{equation}
   \begin{split}
    \vert\Delta\phi_u\vert =& 2\pi\left[\frac{R_G}{a(1 - e^2)} + \frac{6GMR_G}{a^2(1 - e^2)^2}\left(1 + \frac{e^2}{4}\right) \right],
    \label{eq:finalp1}
\end{split} 
\end{equation}
where $a$ denotes {the} orbital semi major axis. This new contribution must be less than the difference between the precession predicted by general relativity and the observed value so that $R_G$ can be inferred using planetary data. Using Mercury's data~\cite{dataset}, we obtain an estimated  value {$|R_G|\lesssim 0.005$ m}. It is important to emphasize that the values calculated  are not constraints of a new theory since we are contrasting our results with derived quantities~\cite{ephemerides,iorio}, calculated from other measured parameters in a particular GR framework. Real constraints should come from calculating all the solar system parameters in the appropriate framework of that~theory.

For other classical GR tests, we  derive the ungravity contributions  using Equation~(\ref{eq:mod0r}). For~light deflection, we take $2K = 0$ in Equation~(\ref{eq:geodesicsK}), as~well as the conserved quantity $h$ in {E}quation ($\ref{eq:conservedqs}$), and in terms of $u = R^{-1}$, we obtain 
\begin{equation}
    u^{\prime\prime} + u = 3GMu^2 - \left(2d_U + 1\right)2GMR_G^{2d_U - 2}u^{2d_U}.
    \label{eq:diffecluz}
\end{equation}
The above equation is solved for $d_U = 3/2$ (as it is our case of interest) using a perturbation method in terms of $GMu/c^2\ll1$. We obtain the deviation from the straight line solution $u = \sin \phi /D$, where $D$ is the closest distance from the light ray trajectory to the gravitational source. In~the limit for large $R$, the~ asymptotic  trajectory and the apparent trajectory form the deflection angle. {Considering that  $R_G\ll D^2$, the deflection angle is }
\begin{equation}
    \delta_U \approx \frac{4GM}{D}\left[1 - \frac{3\pi R_G}{4D}\right].
\end{equation}
We can constrain the parameter $R_G$  {by using the deflection caused by} the Sun~\cite{data1}; this gives the bound {$|R_G|\lesssim1.035\times10^7$ m}. 

The ungravity correction for the gravitational redshift is calculated with the modified  $g_{00}$ term of the metric and the weak field approximation $g_{00} \approx 1 + 2\Phi/c^2$, where $\Phi$ is the classical potential such that

\begin{equation}
    \frac{\Delta_\nu}{\nu} = GM\left[\frac{1}{r_1} - \frac{1}{r_2} - R_G\left(\frac{1}{r_1^2} - \frac{1}{r_2^2}\right)\right],
\end{equation}
where $r_1$ is the radius of the emitter, and $r_2$ is the radius of the detector of a shifted photon. Experimental data from~\cite{data2} lead to the relation { $|R_G|\lesssim3.32\times10^2$ m}.

For completeness, the~ungravity contribution to {the} Shapiro time delay is calculated
\begin{equation}
    d t_U= \pm\left(\frac{2 G M R_G}{r \sqrt{r^2-D^2}}-\frac{D^2 G M R_G}{ r^3 \sqrt{r^2-D^2}}\right) d r,
\end{equation}
and {using data from~\cite{data3}}, the~free parameter is constrained as  {$|R_G| \lesssim 1.1\times10^5$ m}.
\section{Entropic Contributions to the Orbital~Precession}
\label{sec:entropic}
Based on Verlinde's derivation of classical gravity as an entropic force by employing a holographic surface~\cite{verlinde_2011} and considering the thermodynamics relation $F\Delta x = T\Delta S$, one can calculate modifications to Newtonian gravity by adding corrections to the entropy--area relationship. The~modified Newtonian force is given by 
\begin{equation}\label{ver}
    \mathbf{F} = -\left.\frac{GMm}{R^2}\left[1 + 4l_P^2\frac{\partial\mathcal{S}}{\partial A}\right]\right|_{A=4\pi R^2}\hat{R},
\end{equation}
where $\mathcal{S}$ is a modified Bekenstein--Hawking entropy as a function of $A$, the~area of a holographic closed surface between a system formed by a rest mass $M$ and~a test particle $m$. If~a volumetric correction\footnote{In~\cite{isaac}, the~authors study a volumetric correction to the entropy and show that it is the most relevant contribution for galactic rotation curves.} to the entropy is considered
\begin{equation}
    \frac{S\left[A\right]}{k_B} = \frac{A}{4l_P^2} + \epsilon\left(\frac{A}{2l_P^2}\right)^{\frac{3}{2}},
    \label{eq:entropy}
\end{equation}
using Equation~(\ref{ver}), the~modified Newtonian force is
\begin{equation}
\mathbf{F}_{M}=-\frac{GMm}{R^{2}}\left(1+\frac{3\sqrt{2\pi}  }{l_{P}} \epsilon R\right) \hat{R}. \label{modforce}
\end{equation}
As in the usual Kepler problem, the~angular momentum is conserved, and the orbits are restricted to a plane, so we identify $R^2\dot{\phi}=h$ with the magnitude of the angular momentum. Taking 
the radial equation  with  the change of variable $u=R^{-1}$ and after defining $\mu \equiv GM$, we obtain
\begin{equation}
\frac{d^2u}{d\phi^2} + u = \frac{\mu}{h^2}\left(1 + \frac{3\sqrt{2\pi}}{l_P}\epsilon u^{-1}\right).
\end{equation}
Solving perturbatively, we can calculate the perihelion shift.
In terms of the constant free parameter $\epsilon$, the~orbital precession contribution is~\cite{gemma}
\begin{equation}
    \Delta_\phi \approx - \frac{3\pi\sqrt{2\pi}a(1-e^2)}{l_P}\epsilon.
\end{equation}
This extra contribution to the orbital precession must be less than the difference between the observed precession and the {one} predicted by GR. The~bound is calculated using the data for Mercury, resulting in $|\epsilon|\leq 2.8\times 10^{-58}$.

Other modifications to the entropy--area relationship can be considered. One interesting option is the  general entropy presented in~\cite{Nojiri:2022dkr, Odintsov:2023qfj}. This entropy reproduces several generalizations to Shannon's entropy for particular values of the parameters. It has been studied in the context  of cosmology, more precisely, to~understand the dark energy sector. The~generalized entropy  is given by
\begin{equation}
    S_g=\frac{1}{\gamma}\left[\left(1+\frac{\alpha_+}{\beta} S_{BH}\right)^\beta-\left(1+\frac{\alpha_{-}}{\beta} S_{BH}\right)^{-\beta}\right],
    \label{eq:gentropy}
\end{equation}
where $S_{BH}$ is the Bekenstein--Hawking entropy. This reduces to various known entropies (see Table~\ref{tab:entropies}) by fixing the free parameters $\alpha_+$, $\alpha_-$, $\beta$ and $\gamma$, which are constrained to be~positive. 

\begin{table}[h]
\caption{Different 
 entropies can be derived by fixing the positive free parameters in  Equation~(\ref{eq:gentropy}). These are written as a function of the Bekenstein--Hawking entropy $S_{BH}$.}
    \label{tab:entropies}
\small
\setlength{\tabcolsep}{4.08mm}
    \begin{tabular}{ccc}
    \toprule
		\textbf{Entropy} & \textbf{Parameters} & \textbf{Entropy--Area Relationship}\\
	\midrule
		Tsallis--Barrow & $\alpha_+ \rightarrow \infty, \alpha_{-}=0, \gamma=\left(\alpha_{+} / \beta\right)^\beta$ & $S = S_{BH}^\beta$\\
		Rényi &$\alpha_- = 0,\alpha_+=\gamma,\beta \rightarrow 0,\alpha=\frac{\alpha_{+}}{\beta} \rightarrow$ finite &  $S=\frac{1}{\alpha} \ln (1+\alpha S_{BH})$\\
        Sharma--Mittal & $\alpha_{-}=0, \gamma=k=\alpha_{+}, \beta=k / \delta$ &  $S =(1+\delta S_{BH})^{k / \delta}-1$\\
        Kaniadakis & $\beta \rightarrow \infty, \alpha_+=\alpha_-=\gamma/2=K$& $S = \sinh(KS_{BH})$\\
        LQG & $\alpha_-=0, \beta \rightarrow \infty, \gamma=\alpha_+=(1-q)$ & $S=e^{(1-q) S_{BH}}-1$\\
	\bottomrule
\end{tabular}
\end{table}

Following~\cite{verlinde_2011}, the~modified Newtonian force is
\begin{equation}
    \mathbf{F} = -\frac{G M m}{R^2} \frac{1}{\gamma}\left[\alpha_{+}\left(1+\frac{\alpha_{+} \pi}{\beta l_p^2} R^2\right)^{\beta-1} +\alpha_{-}\left(1+\frac{\alpha_{-} \pi R^2}{\beta l_p^2}\right)^{-\beta-1} \right]\hat{R}.
    \label{eq:force0}
\end{equation}
 For planetary orbits, the~modified Binet equation is obtained from the generalized entropy in analogy with the procedure shown in~\cite{gemma}.  Using the conservation of angular momentum  and the change in variable $u =R^{-1}$, we obtain 
\begin{equation}
    u^{\prime \prime}+u=\frac{\mu}{h^2}\left[\frac{\alpha_{+}}{\gamma}\left(1+\frac{\alpha_{+} \pi}{\beta l p^2 u^2}\right)^{\beta-1} +\frac{\alpha_{-}}{\gamma}\left(1+\frac{\alpha_{-} \pi}{\beta l_p^2 u^2}\right)^{-\beta-1} \right],
    \label{eq:modBinetS}
\end{equation}
which can be compared with experimental data by fixing the free parameters.\footnote{We will restrict to $0> \beta> 2$, as~for other values of  $\beta$ we have positive powers of R.}~
It is noticed from Table~\ref{tab:entropies} that Tsallis--Barrow and Sharma--Mittal entropies only recover a Newtonian force term if $\beta = 1$, which is the case for the Bekenstein--Hawking entropy. For~Kaniadakis and LQG entropy, the~limit $\beta \rightarrow \infty$ is used, inconsistently with the values of $\beta$, which lead to an asymptotically null force; the resulting forces only converge if $K=0$ and $q=1$, respectively, and~in this case, the force is reduced to Newton's law.  
For Rényi entropy, the~contribution to the orbital precession is
\begin{equation}
\Delta = 2\pi\left(1 - \frac{\mu^2 l_p^2}{\alpha\pi h^4}\right), 
\end{equation}
and the parameter $\alpha$ is bounded as $\alpha \leq 2.7\times10^{-92}$, comparing with data for~Mercury.\\
Notice that fixing the parameters as $\alpha_- = 0$, $\beta = \frac32$ and $R\longrightarrow\infty$ in Equation ($\ref{eq:force0}$) leads to
\begin{equation}
    F =-\frac{G M m\alpha_+}{\gamma l_p R} \sqrt{\frac{2}{3} \pi \alpha_{+}}.
\end{equation}

This equation can be used to describe stars far from {the} galaxy center, then used to analyze galaxy rotation curves, with~the orbital velocity

\begin{equation}
    v^2 =\frac{G M\alpha_+}{\gamma l_p} \sqrt{\frac23 \pi \alpha_{+}}.
    \label{eq:velocity}
\end{equation}

Comparing with MOND, for~$(\alpha_+)^{3/2}/\gamma \simeq 10^{-56}$, the model is consistent with galactic rotation curves.
Unfortunately, for~small $R$, it is inconsistent with Newton's gravitational~law.

We can also consider that the generalized entropy is a correction of the form $S = S_{BH} + S_g$, then the modified Newtonian gravitational force is
\begin{equation}
    F = -\frac{G M m}{R^2}\left[1 + \frac{\alpha_{+}}{\gamma}\left(1+\frac{\alpha_{+} \pi}{\beta l_p^2} R^2\right)^{\beta-1} +\frac{\alpha_{-}}{\gamma}\left(1+\frac{\alpha_{-} \pi R^2}{\beta l_p^2}\right)^{-\beta-1} \right].
    \label{eq:force1}
\end{equation}

Of particular interest is the behavior for large $R$ 
 and $\beta = \frac32$. In~this case,
\begin{equation}
F=-\frac{G M m}{R^2}\left[1+\frac{\alpha+}{\gamma} \sqrt{\frac{2 \pi \alpha_{+}}{3}} \frac{R}{l_p}+\sqrt{\frac{2 \pi \alpha_{+}}{3}} \frac{3 l_p}{4 \gamma\pi R}\right]+\mathcal{O}\left(\frac{1}{R^5}\right).
\end{equation}
Taking the first correction term in the brackets (which is $R^{-1}$) and with the considerations that lead to Equation~(\ref{eq:modBinetS}), the~differential equation for $u(\phi)$ is
\begin{equation}
    u^{\prime \prime}+u=\frac{\mu}{h^2}\left[1+\frac{\alpha+}{\gamma} \sqrt{\frac{2 \pi \alpha_{+}}{3}} \frac{1}{l_p u}\right].
\end{equation}
In analogy with the procedure described before, this modified Binet equation is solved and the shift of the perihelion, constrained by {the} GR contribution and experimental data, is
\begin{equation}
    |\Delta|=2 \pi\left| \frac{h^2}{\mu} \frac{\alpha_+}{\gamma} \sqrt{\frac{2 \pi \alpha_{+}}{3}} \frac{1}{2 l_p}\right| \leq 2 \pi \times 10^{-12} \frac{\mathrm{rad}}{\mathrm{rev}},
    \label{eq:b32}
\end{equation}
and the entropy free parameters are bounded by $\dfrac{(\alpha_+)^{3/2}}{\gamma} \leq 4.026\times 10^{-58}$. 

If we consider the circular motion of a star far from the galactic center,
the velocity obtained in our model is a constant; 
this is the same behavior one obtains from MOND. After~comparing with  MOND, we obtain
\begin{equation}
\frac{\alpha_+^{3/2}}{\gamma}=\sqrt{\frac{3\hbar a_0}{2\pi c^3M}},
\end{equation}
where $l_p = \sqrt{\hbar G/c^3}$ {and} $a_0 = 1.2\times10^{-10}$ {m/s$^2$}. For~our galaxy, we obtain ${(\alpha_+)^{3/2}}/{\gamma} = 1.12\times10^{-56}$. Comparing with the bounds from the perihelion shift, this model is discarded as an explanation of the anomalous galactic rotation~curve.

\section{Discussion and Final~Remarks}
\label{sec:discussion}
In this paper, we considered the effects of ungravity and entropic gravity on
planetary orbits, with~the goal of establishing the bounds to the parameters of these theories 

Ungravity corrections were previously studied in the cosmological context~\cite{Diaz-Barron:2019uzd}, providing an ungravity origin to the cosmological constant. In~this model, the effective cosmological constant
\begin{equation}
    \Lambda_{e f f} \sim \Lambda_{u}^{2}\left(\frac{M_{u}}{M_{p l}}\right)^{\frac{2}{d_u - 1}},
    \label{eq:cosmo}
\end{equation} 
is given in terms of the ungravity scale $\Lambda_u$, the~ungravity coupling constant $M_u$ and {the} scaling parameter\footnote{In particular, for~$d_u=3/2$ the resulting theory is consistent with an entropy that has a volumetric correction.} $d_u$. {For $d_u=\frac32$, the~effective cosmological constant~\cite{Diaz-Barron:2019uzd} can be written in terms of $R_G$ using Equation ($\ref{eq:Rg}$) as~\begin{equation}
    \Lambda_{eff} \sim \frac{1}{R_G^2}.
\end{equation}
Using the bounds for $R_G$ obtained from different ungravity and unparticle effects, the~value of the effective cosmological constant can be calculated.}

If we assume that the free parameters of ungravity and unparticles are the same (although this is not necessarily true), particle observations can also be considered to bound ungravity parameters.  
Assuming that ungravity and unparticle parameters ($M_u,\Lambda_u,d_u$) have the same values, we can make predictions  using both gravity and particle experiments. Using the unparticle contributions to the hydrogen atom's ground state, we can fix the remaining parameters and calculate the perihelion shift. From~\cite{Wondrak:2016itb}, the~parameters are related as follows
\begin{equation}
    \lambda = c_u\frac{\Lambda_u}{M_u}^k,
\end{equation}
where $\lambda$ is the coupling constant, and $C_u$ together with $k$ are dimensionless constants related to unparticle operators. A~modification term $V_u$ to the potential is added and, using first-order Rayleigh--Schrodinger perturbation theory, the~contribution to {the} ground  state is $\Delta_{100}^{(1)} = \langle 100^0 | V_u | 100^0\rangle$,
and it is related to the  parameters as
\begin{equation}
     \Delta_{100}^{(1)} = -2\frac{\lambda^2\alpha^2\mu^2}{2\Lambda_u^{2d_u - 2}},
\end{equation}
where $\mu$ is the reduced mass, and $\alpha$ is the square of the electron charge.
This new  contribution to the energy can be bounded by  experimental and theoretical errors as $|\Delta_{100}^{(1)}| = \delta E_{th} + \delta E_{exp}$ with the maximum error   $\delta_{max} = (\delta E_{th} + \delta E_{exp})/|E_{th}^s| \approx 1.1\times10^{-5}$. We can write the relation between parameters as
\begin{equation}
    \left|\frac{\Delta_{100}^{(1)}}{E_{th}^s}\right| = \frac{\lambda^2\mu}{\Lambda_u} < \delta_{max},
\end{equation}
and write the orbital precession contribution by introducing $R_G$ in terms of these parameters. The~perihelion shift is
\begin{equation}
    \vert\Delta\phi_u\vert = \frac{2M_{pl}^2}{\pi\Lambda_u^3}\left(\frac{\lambda}{c_u}\right)^{2/k}\left[\frac{1}{a(1 - e^2)} + \frac{6GM}{a^2(1 - e^2)^2}\left(1 + \frac{e^2}{4}\right) \right].
    \label{eq:deltaH}
\end{equation}
We can see that {the} bounds for $\lambda$ and $c_u$ derived from atomic physics will give insignificant orbital precession contributions, emphasizing that unparticle parameters are not necessarily the same as the ungravity~ones. 

From the gravitational classical tests, we obtain the following  bounds for $R_G$: for~light bending  {$|R_G|\lesssim 1.035\times 10^7$ m}, for~the Shapiro time delay  {$R_G \leq 1.1\times10^5$ m}, for  gravitational redshift {$R_G\lesssim 3.32\times10^2$ m} and for precession {$|R_G|\lesssim 0.005$ m}. As~stated before, {these are not constraints} but estimated values for the parameters of the theory.  In~the non-gravitational sector and assuming  that the parameters of ungravity and unparticles are the same,  $\Lambda_u$ and $M_u$ can be calculated from bounds in~\cite{Wondrak:2016itb}. From~these results,  the~value derived for the cosmological constant  is incompatible with the cosmological~observations.
 
Another modification to gravity that we have considered is derived from  modified entropy--area relationships. In~particular, we use a general expression for the entropy that in particular limits the reproduction of several non-additive entropies. This general entropy has been used in the context of cosmology, more precisely in connection to the dark energy sector. 
As in the case of ungravity, we use the perihelion shift  in order to verify the validity of the resulting entropic gravitational force. We find that the contribution to the perihelion shift is negligible. We also study an entropy--area relationship constructed as the sum of the Hawking--Bekenstein entropy and this general entropy.  In~particular, for~$\beta=3/2$ and large $R$, flat rotation curves are predicted. Furthermore, we can fix the value of the remaining parameters by comparing with MOND and obtain ${(\alpha_+)^{3/2}}/{\gamma} = 1.12\times10^{-56}$. Unfortunately, this value is inconsistent with the bounds obtained from the perihelion~shift.

In summary, using the perihelion shift and the solar system data, we can obtain maximal values for the parameters of ungravity as well as for different models of entropic gravity. In~the case of ungravity, we conclude that, with~this methodology, the~bounds on the ungravity parameters are incompatible with the cosmological observations for $\Lambda$, discarding  ungravity  as an origin for the cosmological constant.
For entropic gravity, one can have a modified entropy--area relationship that is consistent with the bounds of dark energy and planetary motion, but~when also considering galactic rotation curves, 
the solar system bound on the parameters favors an interpretation where the volumetric contribution is relevant at the cosmological scale but~not at the galactic~scale.

Finally, we  would like to emphasize that combining solar system and galaxy rotation curve data  is a useful tool to discard modifications to  gravity~\cite{Gaia_DR3,Hees_Folkner,Hees_2015,Desmond_2024,Vokrouhlicky}.



\acknowledgments{The authors would like to thank the anonymous referees who provided valuable comments which helped improve the manuscript. G.P.-C. is supported by the CONAHCyT program  ``Becas Nacionales''. M.S. is supported by grant CIIC 034/2024.}


 \end{document}